\documentclass{article}

\usepackage[english]{babel}

\usepackage[letterpaper,top=2cm,bottom=2cm,left=3cm,right=3cm,marginparwidth=1.75cm]{geometry}

\usepackage{amsmath}
\usepackage{graphicx}
\usepackage[colorlinks=true, allcolors=blue]{hyperref}

\usepackage{amsfonts}
\usepackage{bm}
\usepackage{threeparttable}
\usepackage{diagbox}
\usepackage{makecell}
\usepackage[dvipsnames]{xcolor}
\usepackage{enumitem}
\usepackage{booktabs}
\usepackage{multirow}

\title{Intent-Enhanced Data Augmentation for Sequential Recommendation}
\author{
  Shuai Chen$^{1}$, Zhoujun Li$^{1}$\\
  $^{1}$State Key Laboratory of Complex \& Critical Software Environment,\\
  Beihang University, Beijing 100191, China
}

\begin{document}
\maketitle

\begin{abstract}
The research on intent-enhanced sequential recommendation algorithms focuses on how to better mine dynamic user intent based on user behavior data for sequential recommendation tasks. Various data augmentation methods are widely applied in current sequential recommendation algorithms, effectively enhancing the ability to capture user intent. However, these widely used data augmentation methods often rely on a large amount of random sampling, which can introduce excessive noise into the training data, blur user intent, and thus negatively affect recommendation performance. Additionally, these methods have limited approaches to utilizing augmented data, failing to fully leverage the augmented samples. We propose an intent-enhanced data augmentation method for sequential recommendation(\textbf{IESRec}), which constructs positive and negative samples based on user behavior sequences through intent-segment insertion. On one hand, the generated positive samples are mixed with the original training data, and they are trained together to improve recommendation performance. On the other hand, the generated positive and negative samples are used to build a contrastive loss function, enhancing recommendation performance through self-supervised training. Finally, the main recommendation task is jointly trained with the contrastive learning loss minimization task. Experiments on three real-world datasets validate the effectiveness of our IESRec model.
\end{abstract}

\section{Introduction}
The recommendation system, as a crucial tool for information filtering, has developed over several decades. From early content-based recommendations to today's sequential recommendations, recommendation systems have undergone multiple technological transformations and methodological innovations. The concept of recommendation systems can be traced back to the early 1990s, when recommendation methods were predominantly content-based. This approach recommends items with similar characteristics by analyzing users' preferences for the content features of items (such as text, images, etc.). For example, Lang \cite{lang1995newsweeder} proposed a content-based news recommendation system in his study, which utilized users' preferences for news articles to recommend new content. This method leverages text analysis and feature extraction techniques to capture user interests and provide relevant recommendations, thereby enhancing user satisfaction and engagement.

In the mid-1990s, Collaborative Filtering (CF) methods gradually emerged and became the mainstream in recommendation system research. Collaborative filtering techniques recommend items by analyzing the similarities between users or between items. These methods are mainly divided into user-based collaborative filtering (User-Based CF) and item-based collaborative filtering (Item-Based CF). User-based CF identifies users similar to the target user and recommends items based on the preferences of those similar users, while item-based CF finds items similar to the target item and recommends those items to the user. Resnick et al.'s study \cite{resnick1994grouplens} was one of the early representatives of collaborative filtering, where they proposed the GroupLens system, which utilized users' historical rating data for recommendations. This approach constructs a user-item rating matrix and analyzes the similarity between users’ ratings to make recommendations.

Entering the 21st century, with the enhancement of computational power and the accumulation of large-scale data, matrix factorization (MF) techniques gained widespread attention in recommendation systems. Matrix factorization decomposes the user-item rating matrix into two low-dimensional matrices, representing the latent features of users and items, thus capturing the latent relationships between them. Sarwar et al.'s \cite{sarwar2000application} recommendation algorithm based on singular value decomposition (SVD) significantly improved recommendation accuracy. Koren et al.\cite{koren2009matrix} further advanced matrix factorization techniques by introducing the latent factor model to capture the underlying relationships between users and items. This method not only enhances recommendation accuracy but also exhibits good scalability, making it suitable for large-scale recommendation systems.

\begin{figure}[htbp]
\centering
\includegraphics[width=1.0\columnwidth]{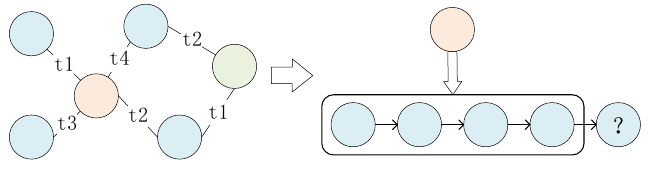}
\caption{Schematic Diagram of Sequential Recommendation from a Graph Perspective}
\label{fig:SRec_intro}
\end{figure}

Sequential recommendation is an important branch of recommendation algorithms, aiming to predict users' future behavior by analyzing their historical behavior sequences. Unlike traditional recommendation algorithms based on static graphs, sequential recommendation focuses on the temporal order and dynamic changes in user behavior. From the perspective of graphs, traditional recommendation algorithms predict the likelihood of a link between a target user and item nodes with which they have not yet interacted, based on existing user-item node interactions. Sequential recommendation algorithms use training data and prediction goals similar to conventional recommendation algorithms, with the key difference being the inclusion of temporal attributes in interactions. As shown in Figure \ref{fig:SRec_intro}, when the interactions between user and item nodes incorporate time attributes, the item nodes associated with the same target user can be arranged into a directed graph in chronological order. The task then becomes predicting which item nodes the user node is likely to be associated with at the next time point.

In summary, the core idea of sequential recommendation is to leverage the temporal information of user behaviors to capture the dynamic changes in user preferences, providing more accurate and personalized recommendations. Compared to other recommendation algorithms, sequential recommendation has the following advantages:
1) Capturing temporal dependencies: Sequential recommendation can capture the temporal dependencies and sequential patterns in user behavior. Unlike traditional collaborative filtering or content-based recommendations, sequential recommendation focuses on the order of user behaviors, enabling the system to better predict users' next actions.
2) Handling dynamic preferences: Users' preferences change over time. Sequential recommendation can more accurately capture users' current interests by analyzing their recent behaviors, thus generating more real-time and dynamic recommendation results.
3) Enhancing recommendation relevance: Since sequential recommendation leverages users' most recent activities, the generated recommendations are typically more aligned with users' current interests. This approach better meets users' immediate needs, improving the accuracy and relevance of recommendations.

The general process of sequential recommendation is described as follows:

1) Data Preprocessing: Collect users' behavior data, including browsing history, click records, purchase history, and rating data. Clean the noisy data and handle missing values to ensure data quality. Arrange the users' behavior data in chronological order and split them into training, validation, and test sets.
2) Constructing Sequential Data: Organize users' behavior data into sequences according to their time order, where each sequence represents a user's continuous behavior. For each subsequence, generate corresponding target labels, i.e., predicting the user's next behavior after the current behavior sequence.
3) Sequence Modeling: Choose an appropriate sequence model architecture, such as Recurrent Neural Networks (RNN) \cite{elman1990finding}, Long Short-Term Memory Networks (LSTM) \cite{hochreiter1997long}, or Transformer. Design the model structure, including input, hidden, and output layers, and determine model hyperparameters (such as the number of layers, hidden units, activation functions, etc.).
4) Model Training: Convert the users' behavior sequences into feature vectors and input them into the model. Train the model using the training dataset to optimize the model parameters. Define and compute the loss function (such as cross-entropy loss, mean squared error, etc.) and update the model parameters using the backpropagation algorithm.
5) Generating Sequential Recommendations: Use the trained model to predict the next possible behavior or item based on the user's current behavior sequence. Based on the prediction results, generate a recommendation list and recommend the highest-scoring items to the user.

In academia, Hidasi et al. proposed a sequential recommendation model, GRU4Rec \cite{DBLP:journals/corr/HidasiKBT15}, based on the Gated Recurrent Unit (GRU), which improves recommendation accuracy by capturing the temporal dependencies of user behavior. This model utilizes GRU to model users' session data, thereby capturing the short-term dynamic changes in users' interests. The SASRec model \cite{kang2018self} further enhances recommendation performance by employing the self-attention mechanism for sequential recommendation. SASRec captures long-range dependencies in user behavior sequences through the self-attention mechanism, enabling more accurate predictions of users' future actions. CL4SRec \cite{xie2022contrastive} uses data augmentation techniques such as item cropping, insertion, and reordering, and leverages contrastive learning principles to enhance sequential recommendation performance. CL4SRec effectively handles noise during the data augmentation process through contrastive learning, thus improving the robustness and accuracy of recommendations. ICLRec \cite{chen2022intent} introduces latent intents in users' behavior sequences into the sequential recommendation model, using contrastive self-supervised learning (SSL) to maximize the consistency between the sequence view and its corresponding intents, thereby enhancing recommendation performance and model robustness. DuoRec \cite{qiu2022contrastive} proposes a contrastive learning method to address representation degradation in sequential recommendation, thereby improving the performance of the recommendation system. DCRec \cite{yang2023debiased} introduces a new debiased contrastive learning paradigm, combining sequential pattern encoding with global collaborative relationship modeling through an adaptive compliance-aware augmentation method. This approach aims to resolve the popularity bias issue in recommendation systems and effectively capture intra-sequence item transition patterns and cross-sequence user dependencies, significantly enhancing recommendation performance. MAERec \cite{ye2023graph} proposes a simple and effective graph-masked autoencoder-enhanced sequential recommendation system. By adaptively and dynamically extracting global item transition information for self-supervised augmentation, it avoids the dependence on high-quality contrastive views in existing contrastive learning models. This approach addresses the challenges of label scarcity and noise in user behavior data, thereby significantly improving the representation capability and recommendation performance in sequential recommendation tasks.

Various data augmentation methods have been widely applied in sequential recommendation algorithms. As a precursor to contrastive learning, many models based on contrastive learning also strongly rely on different data augmentation methods. Some studies have pointed out that merely applying data augmentation, without incorporating contrastive loss functions to aid training, can achieve similar recommendation performance to joint training with contrastive loss \cite{zhou2024contrastive}. Common data augmentation methods used in sequential recommendation tasks can be divided into two categories: the first type does not alter the training data labels, with examples including random deletion, insertion, and masking of nodes; the second type changes the training data labels, with the sliding window data augmentation as a typical example. The sliding window data augmentation method has been widely used in various sequential recommendation algorithms, yielding excellent results.

However, these methods have a common drawback, which is their reliance on randomness, either introducing noise into the user intent sequence or diminishing the user intent. This may lead to several issues: 
1) Increased noise in the training data: Random insertion or deletion of nodes may introduce irrelevant information, which could cause the model to mistakenly treat this noise as meaningful signals, thus affecting recommendation performance. This noise interferes with the learning process, making it difficult for the model to accurately capture users' true intentions.
2) Blurring of user intent: Random operations may disrupt the inherent structure of the user behavior sequence, making the user's intent within the sequence unclear. This hinders the model’s ability to extract effective features, thereby impacting the accuracy of the recommendation results.
3) Lack of semantic information: Simple randomness-based data augmentation fails to fully exploit the semantic information in user behavior sequences, leading to the model's inability to effectively understand and utilize this information. As a result, the model may struggle to make accurate recommendations when faced with complex user behavior patterns.

To address the challenges in sequential recommendation, we propose an Intent-Enhanced Sequence Recommendation Algorithm (IESRec). The core idea of the IESRec model is to leverage intent insertion to construct positive and negative samples that incorporate user intent for the sequential recommendation task. First, we generate an artificial sequence of a fixed length that aligns with the user's historical behavior and insert it into a non-terminal position in the training data to create positive samples. Second, we generate an intent-extended sequence at the terminal position of the training data, ensuring that the extended sequence does not match the original training labels, thereby creating negative samples. The generated positive and negative samples are used to expand the training data and construct a joint contrastive loss training. The contributions of our IESRec model are as follows:

\begin{itemize}[label=$\bullet$]
\item We propose a data augmentation method guided by intent enhancement. Unlike conventional data augmentation methods that heavily rely on randomness, our approach is more effective in accurately extracting users' complex and dynamic intents, thereby improving the performance of the model in sequential recommendation tasks.
\item Through contrastive learning, our model’s robustness is enhanced, ensuring that it can still provide high-quality recommendations even when users' intents change frequently. Especially in dynamic and uncertain environments, our model quickly adapts to shifts in user demands, delivering more personalized recommendation services.
\item Our approach not only introduces a new data augmentation method tailored for sequential recommendation in theory but also demonstrates its effectiveness in experiments. Experimental results show that our model performs exceptionally well in various recommendation scenarios, significantly outperforming existing sequential recommendation methods. The results indicate that the IESRec model exhibits outstanding performance and stability in multiple real-world recommendation settings, highlighting its broad application potential for enhancing overall recommendation system effectiveness across different fields.
\end{itemize}

In conclusion, our IESRec model offers new insights and approaches for improving sequential recommendation tasks, which not only hold significant academic value but also provide strong support for recommendation systems in real-world applications.

\section{Methodology}
In this section, we first introduce the task definition of sequential recommendation, followed by the construction of our IESRec model, and finally, we explain how the model performs joint training and prediction.

\subsection{Task Definition}
This chapter focuses on the sequential recommendation task, defined as follows: Given a user-item interaction graph \(\mathcal{G} = (\mathcal{V}, \mathcal{E})\), where \(\mathcal{V}\) represents the set of all nodes, and \(\mathcal{E}\) represents the set of all edges. The set of user and item nodes in \(\mathcal{V}\) are denoted as \( \mathcal{U} \) and \( \mathcal{I} \), respectively. Each user \( u \in \mathcal{U} \) has a chronologically ordered sequence of interacted items \( I^u = [i^u_1, \ldots, i^u_t, \ldots, i^u_{|I^u|}] \), where \( |I^u| \) represents the number of item nodes interacted with by user \( u \), and \( i^u_t \in \mathcal{I}\) indicates the item interacted with by user \( u \) at time \( t \). For each user \( u \), the task of sequential recommendation is to predict the list of items that the user \( u \) is most likely to interact with at the next time point based on the historical interaction sequence \( I^u \). The top k items from the ranked list are then selected, and metrics such as recall or NDCG are calculated to evaluate the quality of the ranking results.

\subsection{Model Framework}
\begin{figure}[htbp]
\centering
\includegraphics[width=1.0\columnwidth]{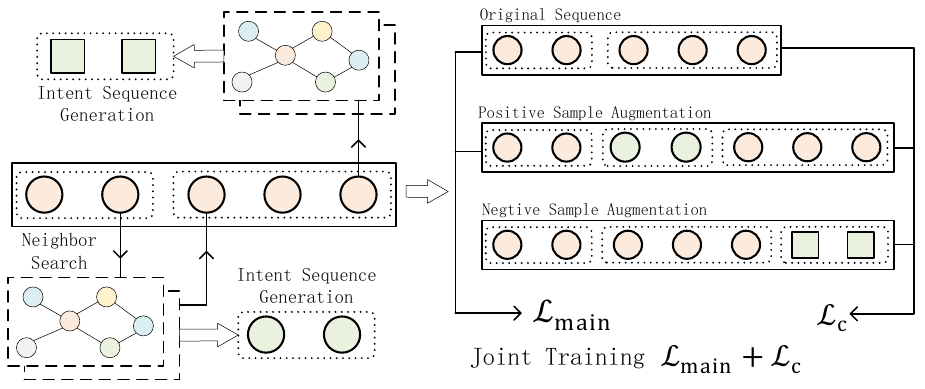}
\caption{The overall architecture of IESRec model}
\label{fig:SRec_model}
\end{figure}

The framework of our IESRec model is shown in Figure \ref{fig:SRec_model}. By generating intent sequences at different positions in the original sequence and inserting them into the corresponding locations, we accomplish the enhancement of positive and negative samples. These enhanced samples, along with the original samples, are jointly trained to improve the model's sequential recommendation performance. Next, we will introduce our model in detail from three aspects: intent insertion data augmentation, behavior sequence encoding, and contrastive learning training.
    
    \subsubsection{Intent Insertion Data Augmentation}

In this section, we introduce how to apply the intent insertion method for data augmentation suitable for sequential recommendation tasks, including both positive and negative sample generation. First, our data augmentation is based on the sliding window data augmentation technique. For simplicity, all user behavior sequences mentioned below are the result of sliding window processing. In the sequential recommendation task, the behavior sequence fragments of users reflect their intents. Unlike most data augmentation methods based on random sampling, our intent insertion data augmentation method generates positive and negative samples by inserting artificial user behavior sequences at different positions within the sequence data.

For the generation of intent-inserted positive samples, given the current user \( u \)'s behavior sequence \( I^u = [i^u_1, i^u_2, \ldots, i^u_T] \), for each adjacent node pair \([i^u_t, i^u_{t+1}], t=1, 2, \dots, T-1\), we attempt to insert an artificial sequence of length \(K\), \([\bar{i}^u_1, \bar{i}^u_2, \cdots, \bar{i}^u_K]\), between the two nodes. The artificial sequence is constructed as follows: First, for each item node, we traverse all the data in the training set and define the set of all nodes appearing after the target node as the adjacency set of that item node. Second, we randomly select an element from the adjacency set of node \(i^u_t\) as the value of \(\bar{i}^u_1\), and continue this process iteratively until the value of \(\bar{i}^u_K\) is determined. At this point, if \(i^u_{t+1}\) belongs to the adjacency set \(\bar{I}^u_k\) of \(\bar{i}^u_K\), then the artificial sequence meets the requirements and is inserted between \([i^u_t, i^u_{t+1}]\) as an intent-inserted enhanced positive sample. The enhanced sample is as follows:

\begin{equation}
\bar{I}^u = [i^u_1, \cdots, i^u_t, \bar{i}^u_1, \cdots, \bar{i}^u_K, i^u_{t+1}, \cdots, i^u_T]
\end{equation}
    
The sequence \([\bar{i}^u_1, \bar{i}^u_2, \cdots, \bar{i}^u_K]\) is the inserted artificial sequence, whose start and end are dependent on the characteristics of the nodes in the original sequence. It serves as a dynamic extension and reinforcement of the user intent reflected in the original sequence. This positive sample augmentation will be mixed into the training set for direct training and will also be used alongside the generated negative samples for contrastive learning.

For the generation of intent-inserted negative samples, given the behavior sequence \( I^u = [i^u_1, i^u_2, \ldots, i^u_T] \), we insert an artificial subsequence of length \(K\) at the end of the sequence for data augmentation. Since the insertion is at the end of the original sequence, this effectively acts as a prediction of the user's intent and an extension of the behavior sequence. As a result, the labels in the training set no longer align with the user intent reflected by the extended behavior sequence, and this data augmentation is considered as generating negative samples. Specifically, we use a method similar to the one for constructing positive samples. Based on the last node \(i^u_T\) and its adjacency set, we construct a sequence of length \(K\), \([\bar{i}^u_1, \bar{i}^u_2, \cdots, \bar{i}^u_K]\). Unlike positive sample construction, negative sample construction does not require that the adjacency set of the last node includes the next node from the original sequence. Therefore, the augmented negative sample we obtain is:

\begin{equation}
\hat{I}^u = [i^u_1, \cdots, i^u_T, \hat{i}^u_1, \cdots, \hat{i}^u_K]
\end{equation}

The sequence \([\hat{i}^u_1, \hat{i}^u_2, \cdots, \hat{i}^u_K]\) is the artificial sequence inserted at the end. This inserted sequence serves as a prediction of the user's dynamic intent, and the enhanced sample no longer matches the training label nodes of the original sample. This negative sample will be used alongside the positive samples to construct the contrastive loss function for auxiliary training.
    
\subsubsection{Behavior Sequence Encoding}

In our IESRec model, we use a Transformer encoder to encode each item sequence. First, the user's behavior sequence \( I^u \) is transformed into an embedding representation \( \mathbf{S}^u = [\mathbf{s}^u_1, \ldots, \mathbf{s}^u_t, \ldots, \mathbf{s}^u_T] \), where \( \mathbf{s}^u_t \) is the \( d \)-dimensional embedding vector of item \( s^u_t \). To preserve the temporal order information of the item sequence, we construct a positional encoding matrix \( \mathbf{P} \in \mathbb{R}^{T \times d} \), where \( T \) represents the maximum length of all sequences. The item embeddings and positional encodings are summed to form the input vector at time point \( t \) for the Transformer:
    \begin{equation}
    \mathbf{h}_t^0 = \mathbf{s}_t + \mathbf{p}_t
    \end{equation}
    
Here, \( \mathbf{h}_t^0 \in \mathbb{R}^d \) is the initial input vector at time point \( t \), and \( \mathbf{p}_t \) is the positional encoding at time \( t \). Subsequently, we use the multi-head attention mechanism to compute the representation of each item node. Assuming \( \mathbf{H}^0 = [\mathbf{h}_0^0, \mathbf{h}_1^0, \ldots, \mathbf{h}_t^0] \) is the initial vector representation of the sequence, we input it into a multi-head Transformer encoder (Trm) with \( L \) layers:
    
    \begin{equation}
    \mathbf{H}^L = \text{Trm}(\mathbf{H}^0)
    \end{equation}
    
The final hidden vector \(\mathbf{h}^L_t\) in the set of hidden vectors from the last layer \( \mathbf{H}^L = [\mathbf{h}_0^L, \mathbf{h}_1^L, \ldots, \mathbf{h}_t^L] \) is selected as the final representation of the user behavior sequence \( S^u \). For convenience in subsequent usage, we denote it as \(\mathbf{h}_u\).
    
\subsubsection{Contrastive Learning Training}

In this section, we introduce how contrastive learning enhances the model's ability to interpret and predict the user's dynamic intent within behavior sequences. Our core objective is to enable the model to better identify the user's current intent when processing behavior sequences, thus making more accurate predictions. To achieve this, we adopt the contrastive learning paradigm, utilizing the augmented user behavior sequences, including both positive and negative samples, to improve the model's robustness and predictive capability.

Specifically, the positive sample pairs for contrastive learning consist of the original sequence \( I^u \) and its corresponding intent-inserted positive augmented sequence \( \bar{I}^u \), where we expect the model's prediction results for these two sequences to be as similar as possible. On the other hand, the negative sample pairs consist of the original sequence \( I^u \) and its corresponding negative augmented sequence \( \hat{I}^u \), where we expect the model's predictions for these two sequences to be as different as possible. We then use a contrastive loss function to train the model, maximizing the similarity of positive sample pairs and minimizing the similarity of negative sample pairs. The contrastive loss function is defined as follows:
   \begin{equation}
   \mathcal{L}_c = -\log \frac{\exp(\text{sim}(\mathbf{h}_u, \bar{\mathbf{h}}_u) / \tau)}{\exp(\text{sim}(\mathbf{h}_u, \hat{\mathbf{h}}_u) / \tau)}
   \end{equation}
   
Here, \(\mathbf{h}_u\), \(\bar{\mathbf{h}}_u\), and \(\hat{\mathbf{h}}_u\) represent the representations of the original sequence and its corresponding positive and negative sample sequences, respectively. \(\text{sim}(\cdot, \cdot)\) denotes the similarity measure, and in our experiments, we use the inner product of vectors as the similarity metric. \(\tau\) is the temperature parameter that controls the intensity of contrastive learning. Through the contrastive learning paradigm, we enhance the model's recommendation performance and robustness when dealing with highly complex user intent in behavior sequences.

\subsection{Joint Training and Prediction}

The model's final loss function combines the loss from the main recommendation task and the contrastive learning loss. The loss for the main recommendation task consists of the model's loss on the original training set and the loss on the positive samples, as shown below:
   \begin{equation}
   \mathcal{L} = \mathcal{L}_{\text{rec}} + \bar{\mathcal{L}}_{\text{rec}} +\lambda \mathcal{L}_{\text{c}}
   \end{equation}
        
Here, \(\mathcal{L}_{\text{rec}}\) and \(\bar{\mathcal{L}}_{\text{rec}}\) represent the cross-entropy losses on the original training set and the generated positive sample data, respectively. We use the hyperparameter \(\lambda\) to adjust the weight of the contrastive learning loss. In the subsequent experiments, we will analyze the effects of \(\lambda\) and \(\tau\) on the model's recommendation performance. During prediction, we do not consider the contrastive learning module or positive sample data augmentation; instead, we directly use the trained model to predict the user's next behavior based on the encoded behavior sequence.

\section{Experiments}
In this section, we first describe our experimental setup, including an introduction to the datasets, baseline models, evaluation metrics for model performance, and training details. We then provide an overall comparison of the performance of our model against the baseline models, followed by ablation experiments for our model. Finally, we conduct a sensitivity analysis of the model's hyperparameters.

\subsection{Experimental Setup}
In this subsection, we first introduce the datasets used for all the experiments in this chapter. We then introduce the baseline models against which we compare our model, categorizing these baseline models. The evaluation metrics refer to the criteria used in our experiments to assess the performance of the sequential recommendation task, following the standard conventions in academia for this task. Lastly, we outline the detailed configurations used for model training.

\subsubsection{Dataset Introduction}
We conduct experiments on three subsets of the widely-used Amazon dataset \cite{mcauley2015image}. The Amazon dataset contains user review data from the Amazon website from 1996 to 2014. The dataset includes information such as user IDs, product IDs, ratings, review text, and timestamps. This dataset is extensively used in the study of recommendation systems, text mining, and sentiment analysis. Specifically, we conduct experiments on three subsets: Beauty, Clothing, and Sports.

The Beauty subset contains user reviews and ratings related to beauty and personal care products. This subset includes a wide range of beauty products, such as skincare, cosmetics, and perfumes. The users in this subset provide detailed ratings and reviews of beauty products, focusing on aspects like product efficacy, packaging, and scent. Although the product variety is extensive, the dataset is relatively sparse, as user purchasing and reviewing behavior in the beauty category is relatively scattered.

The Clothing subset contains user reviews and ratings related to clothing, footwear, and accessories. This subset covers various types of apparel, including men's and women's clothing, sportswear, shoes, and accessories. Users in this subset typically provide reviews about size, comfort, style, and quality. Due to the fast product turnover and strong seasonality, users' purchasing and reviewing behavior exhibits time sensitivity.

The Sports subset contains user reviews and ratings related to sporting goods and outdoor equipment. This subset includes equipment for various sports and outdoor activities, such as sports gear, fitness equipment, and outdoor gear. Users in this subset often review the functionality, durability, and performance of the products, covering a wide range of sports and outdoor activity scenarios.

We remove any user and item nodes that appear less than five times in the dataset. The details of the three datasets are presented in Table \ref{tab:SRec_dataset}, including the number of users and items, the number of user-item interactions, the average length of user behavior sequences, and the density of each dataset.

\begin{table}[htbp]
  \centering
  \caption{Statistics of the Datasets Used in IESRec Model Experiments}
    \begin{tabular}{cccc}
    \toprule
    Dataset   & Beauty & Sports & Clothing \\
    \midrule
    \#User  & 22363 & 35598 & 39387 \\
    \#Item  & 12101 & 18357 & 23033 \\
    \#Interaction  & 198502 & 296337 & 278677 \\
    Average length  & 8.9   & 8.3   & 7.1 \\
    Density   & 0.07\% & 0.05\% & 0.03\% \\
    \bottomrule
    \end{tabular}%
  \label{tab:SRec_dataset}%
\end{table}%

\subsubsection{Baseline Models}  
We compare our model with several classical works in the field of sequential recommendation. These baseline models are divided into two categories: the first category consists of methods that do not use the contrastive learning paradigm, and the second category includes methods that utilize contrastive learning. The details of these baseline models are as follows:

\begin{itemize}[label=$\bullet$]
        
\item \textbf{GRU4Rec\cite{DBLP:journals/corr/HidasiKBT15}} was the first to apply Gated Recurrent Neural Networks (GRU) to sequential recommendation tasks.

\item \textbf{BERT4Rec\cite{sun2019bert4rec}} models data for sequential recommendation tasks by adopting the masked node paradigm, inspired by natural language processing tasks.

\item \textbf{SASRec\cite{kang2018self}} was the first to use a unidirectional self-attention mechanism to model sequential recommendation tasks.

\item \textbf{CL4Rec\cite{xie2022contrastive}} uses cropping, masking, or reordering as data augmentation techniques, and then applies the contrastive learning paradigm for modeling sequential recommendation tasks.

\item \textbf{ICLRec\cite{chen2022intent}} enhances recommendation performance and model robustness by introducing latent intents in user behavior sequences into the sequential recommendation model, using contrastive self-supervised learning (SSL) to maximize the consistency between sequence views and their corresponding intents.

\item \textbf{DuoRec\cite{qiu2022contrastive}} proposes a contrastive learning method to address representation degradation in sequential recommendation, thus improving the performance of the recommendation system.

\item \textbf{DCRec\cite{yang2023debiased}} introduces a new debiased contrastive learning paradigm, which combines sequential pattern encoding with global collaborative relationship modeling through an adaptive compliance-aware augmentation method. This approach aims to resolve popularity bias issues in recommendation systems and effectively captures item transition patterns within sequences and cross-sequence user dependencies, significantly enhancing recommendation performance.

\item \textbf{MAERec\cite{ye2023graph}} proposes a simple and effective graph-masked autoencoder-enhanced sequential recommendation system. By adaptively and dynamically extracting global item transition information for self-supervised augmentation, it avoids the dependence on high-quality embedding contrastive views present in existing contrastive learning models, addressing issues like label scarcity and noise in user behavior data, thereby significantly improving representation capability and recommendation performance in sequential recommendation tasks.

\end{itemize}

\subsubsection{Evaluation Metrics}

We use two commonly used evaluation metrics to assess the recommendation performance of our IESRec model and the baseline models: Hit Rate (HR) \cite{goldberg1992using} and Normalized Discounted Cumulative Gain (NDCG) \cite{jarvelin2002cumulated}. HR is a simple and intuitive metric that measures whether the recommendation system can include the items that users are genuinely interested in within the recommendation list. It represents the proportion of test users for whom the system successfully recommends at least one item of interest. The formula for HR is as follows:

\begin{equation}
    \text{HR} = \frac{\sum_{u \in U} I(\text{rank}(u) \leq k)}{|U|}
\end{equation}

Here, \(U\) is the set of all users. \(I(\cdot)\) is an indicator function that returns 1 when the condition inside the parentheses is true, and 0 otherwise. \(rank(u)\) is the position of the relevant item for user \(u\) in the recommendation list. \(k\) is the length of the recommendation list, and \(|U|\) is the total number of users. NDCG is a ranking-aware evaluation metric used to measure the quality of the recommendation list, with higher scores indicating that the items users are genuinely interested in appear earlier in the list. We adopt the full-ranking strategy \cite{krichene2020sampled}, where the match scores between the given user nodes and all item nodes are ranked from highest to lowest to evaluate the model's recommendation performance. In our experiments, the recommendation list length is set to \(K=10, 20, 50\), meaning we use HR@10, HR@20, HR@50, and NDCG@10, NDCG@20, NDCG@50 to test the models.

\subsubsection{Training Details}

We use the recently released self-supervised learning framework SSLRec \cite{ren2024sslrec} for our experiments. For models not included in this framework, we use the official code and configurations for training. During the data preparation phase, we process the raw data uniformly and fix the training, validation, and test sets, ensuring that all models are compared using the same training and test data. The behavior data of all users is sorted chronologically, with the last 50 behavior records used as the test set. Skipping the last behavior, the previous 50 records (starting from the second to last) are used as the validation set, and the preceding 50 records (starting from the third to last) are used as the training set. The maximum sequence length is fixed at 50, with any length exceeding this being truncated, and shorter sequences are padded with extra placeholders.

During training, the learning rate is uniformly set to 0.001, and the batch size is set to 512. The embedding dimension for all nodes is fixed at 64. The models are trained using the Adam optimizer \cite{DBLP:journals/corr/KingmaB14}, with all dropout values set to 0.5. We apply an early stopping mechanism to determine when to stop training, specifically monitoring the MRR metric on the validation set after each training epoch. If there is no improvement after 10 consecutive checks, training is stopped. 

In terms of hardware, all model training and inference tasks were conducted on an Intel i7-13700K CPU and an NVIDIA TITAN RTX GPU.

\subsection{Overall Performance Comparison}

In our research on recommendation algorithms from a graph perspective, we conducted a comprehensive comparison between our IESRec model and a series of baseline models. This comparison focused on the Beauty, Clothing, and Sports datasets, evaluating their performance based on two metrics: Hit Rate (HR) and Normalized Discounted Cumulative Gain (NDCG). The data is presented in Table \ref{tab:SRec_performance}. The results indicate that our model significantly outperforms the baseline models across all datasets, with a detailed analysis provided below.

\begin{table}[htbp]
  \centering
  \caption{Recommendation Performance of the IESRec Model and Baseline Models on Different Datasets}
  \resizebox{1.0\textwidth }{!}{
    \begin{tabular}{c|c|ccc|ccc|ccc|}
    \toprule
    \multicolumn{2}{c|}{Dataset} & \multicolumn{3}{c|}{Beauty} & \multicolumn{3}{c|}{Sports} & \multicolumn{3}{c}{Clothing} \\
    \midrule
    \midrule
    Model    & K=    & 10    & 20    & 50    & 10    & 20    & 50    & 10    & 20    & \multicolumn{1}{c}{50} \\
    \multirow{2}[0]{*}{GRU4Rec} & HR@K & 0.0309  & 0.0440  & 0.0659  & 0.0259  & 0.0389  & 0.0647  & 0.0150  & 0.0232  & \multicolumn{1}{c}{0.0377 } \\
          & NDCG@K & 0.0180  & 0.0213  & 0.0250  & 0.0141  & 0.0174  & 0.0224  & 0.0079  & 0.0099  & \multicolumn{1}{c}{0.0129 } \\
    \multirow{2}[0]{*}{BERT4Rec} & HR@K & 0.0536  & 0.0764  & 0.1144  & 0.0288  & 0.0433  & 0.0719  & 0.0167  & 0.0258  & \multicolumn{1}{c}{0.0419 } \\
          & NDCG@K & 0.0312  & 0.0369  & 0.0433  & 0.0156  & 0.0193  & 0.0249  & 0.0088  & 0.0110  & \multicolumn{1}{c}{0.0143 } \\
    \multirow{2}[0]{*}{SASRec} & HR@K & 0.0471  & 0.0743  & 0.1283  & 0.0314  & 0.0511  & 0.0870  & 0.0206  & 0.0311  & \multicolumn{1}{c}{0.0514 } \\
          & NDCG@K & 0.0247  & 0.0305  & 0.0408  & 0.0145  & 0.0206  & 0.0275  & 0.0095  & 0.0120  & \multicolumn{1}{c}{0.0150 } \\
    \multirow{2}[0]{*}{CL4SRec} & HR@K & 0.0521  & 0.0809  & 0.1378  & 0.0327  & 0.0514  & 0.0884  & 0.0210  & 0.0326  & \multicolumn{1}{c}{0.0532 } \\
          & NDCG@K & 0.0262  & 0.0334  & 0.0446  & 0.0167  & 0.0215  & 0.0288  & 0.0097  & 0.0126  & \multicolumn{1}{c}{0.0167 } \\
    \multirow{2}[0]{*}{ICLRec} & HR@K & 0.0659  & 0.0952  & 0.1464  & 0.0377  & 0.0564  & 0.0923  & 0.0214  & 0.0318  & \multicolumn{1}{c}{0.0523 } \\
          & NDCG@K & 0.0363  & 0.0437  & 0.0538  & 0.0183  & 0.0225  & 0.0289  & 0.0116  & 0.0142  & \multicolumn{1}{c}{0.0182 } \\
    \multirow{2}[0]{*}{DuoRec} & HR@K & 0.0690  & 0.0967  & 0.1411  & 0.0390  & 0.0579  & 0.0971  & 0.0210  & 0.0324  & \multicolumn{1}{c}{0.0557 } \\
          & NDCG@K & 0.0372  & 0.0434  & 0.0513  & 0.0205  & 0.0249  & 0.0322  & 0.0113  & 0.0142  & \multicolumn{1}{c}{0.0188 } \\
    \multirow{2}[0]{*}{DCRec} & HR@K & 0.0787  & 0.1085  & 0.1600  & 0.0443  & 0.0641  & 0.1001  & 0.0234  & 0.0322  & \multicolumn{1}{c}{0.0591 } \\
          & HR@K & 0.0408  & 0.0471  & 0.0555  & 0.0216  & 0.0258  & 0.0317  & 0.0136  & 0.0158  & \multicolumn{1}{c}{0.0191 } \\
    \multirow{2}[0]{*}{MAERec} & NDCG@K & 0.0794& 0.1160& 0.1787& 0.0452& 0.0686& 0.1100  & 0.0251& 0.0391& \multicolumn{1}{c}{0.0614} \\
          & HR@K & 0.0413& 0.0489& 0.0603& 0.0213& 0.0268& 0.0338  & 0.0126  & 0.0156  & \multicolumn{1}{c}{0.0204 } \\
    \multirow{2}[1]{*}{IESRec}& NDCG@K & \textbf{0.0853 } & \textbf{0.1208 } & \textbf{0.1818 } & \textbf{0.0486 } & \textbf{0.0705 } & \textbf{0.1112 } & \textbf{0.0296 } & \textbf{0.0423 } & \multicolumn{1}{c}{\textbf{0.0647 }} \\
          & HR@K & \textbf{0.0430 } & \textbf{0.0519 } & \textbf{0.0640 } & \textbf{0.0231 } & \textbf{0.0287 } & \textbf{0.0366 } & \textbf{0.0138 } & \textbf{0.0170 } & \multicolumn{1}{c}{\textbf{0.0214 }} \\
    \bottomrule
    \end{tabular}%
  \label{tab:SRec_performance}}%
\end{table}%

\begin{itemize}[label=$\bullet$]
    \item Dataset-specific analysis: In the Beauty dataset, our IESRec model performed exceptionally well in both HR and NDCG metrics. Specifically, at \(K=50\), IESRec achieved an HR of 0.1818, significantly higher than the best-performing baseline model MAERec (HR of 0.1787). For the NDCG metric, our model scored 0.0640, also notably higher than the second-best model MAERec (NDCG of 0.0603). These results indicate that our model has a higher hit rate and ranking quality for beauty product recommendations, allowing it to more accurately capture user preferences. In the Clothing dataset, our model similarly excelled, achieving an HR of 0.0647 at \(K=50\), significantly outperforming the best baseline model MAERec (HR of 0.0614). For the NDCG metric, our model scored 0.0214, surpassing all baseline models. In the Sports dataset, our model showed outstanding performance. IESRec achieved HR and NDCG scores of 0.1112 and 0.0366 at \(K=50\), respectively, significantly surpassing all baseline models. In contrast, the best-performing baseline model MAERec scored 0.1100 in HR and 0.0338 in NDCG. Our model demonstrated higher precision and user satisfaction in recommending sports and outdoor equipment.
    
    \item Analysis of baseline model performance: GRU4Rec showed relatively basic performance, with HR and NDCG scores consistently lower across all datasets, indicating that traditional recurrent neural network methods have limitations in handling complex user-item relationships. BERT4Rec performed better than GRU4Rec across datasets but still did not achieve the best results in either HR or NDCG, indicating improvements in capturing contextual information but still falling short. SASRec, which uses Transformers for sequential recommendation, showed moderate performance, with some improvement in capturing user behavior sequence patterns but still not achieving the best results. CL4SRec performed well across datasets, especially in the Sports dataset, showing an advantage in capturing sequence dependencies. ICLRec performed well in the Beauty dataset but underperformed in other datasets. DuoRec showed stable performance across all datasets but still lagged behind the IESRec model. DCRec and MAERec performed closely to our model in some cases but were still slightly inferior overall.
    
    \item Overall conclusion: The IESRec model outperformed all baseline models across all datasets, achieving significantly higher scores in both HR and NDCG. This indicates that our model has higher accuracy and user satisfaction in handling recommendation tasks across different domains and can better capture and understand users' personalized needs. By inserting artificial intent sequences at different positions in the user's historical behavior sequence, we generated a series of positive and negative samples. Through the effective utilization of these samples, we enhanced the model's ability to recognize and predict users' current intent, demonstrating powerful advantages and broad application prospects in sequential recommendation tasks.
\end{itemize}
    
\subsection{Ablation Study}

In this section, we evaluate the impact of several key components of the model on recommendation accuracy. Specifically, we gradually remove parts of the designed model architecture and observe the changes in accuracy across the three datasets. This approach allows us to quantitatively assess the contribution of our model design to the recommendation performance. Based on modifications to the model structure, we generate two variants of our IESRec model:

\begin{itemize}[label=$\bullet$]
\item w/o PS (without positive samples in training data): This variant removes the positive sample data generated by intent insertion from the training data (while retaining sliding window data augmentation). The contrastive learning module is still preserved.
\item w/o CL (without contrastive learning module): This variant removes the contrastive learning module. In this case, the model is trained only using the sliding window-augmented training samples and the positive samples generated by intent insertion.
\end{itemize}

Table \ref{tab:SRec_ablation} presents the experimental results. By comparing these variants, we can effectively assess the contribution of the positive and negative samples generated through intent insertion to the overall performance of our model.

\begin{table}[htbp]
  \centering
  \caption{Ablation Study Results of the IESRec Model}
    \begin{tabular}{ccccccc}
    \toprule
    \multirow{2}[4]{*}{Dataset} & \multicolumn{2}{c}{\textbf{Beauty}} & \multicolumn{2}{c}{\textbf{Sports}} & \multicolumn{2}{c}{\textbf{Clothing}} \\
\cmidrule{2-7}          & Recall@20 & NDCG@20 & Recall@20 & NDCG@20 & Recall@20 & NDCG@20 \\
    \midrule
    IESRec & 0.1208  & 0.0519  & 0.0705  & 0.0287  & 0.0423  & 0.0170  \\
    w/o PS & 0.1159& 0.0496& 0.0668& 0.0264& 0.0404& 0.0155\\
    w/o CL & 0.1170& 0.0503& 0.0642& 0.0260& 0.0398& 0.0150\\
    \bottomrule
    \end{tabular}%
  \label{tab:SRec_ablation}%
\end{table}%

First, we observe that removing either the generated positive samples from the training data or the contrastive learning module results in a significant decrease in recommendation accuracy. This demonstrates the critical importance of both components for model performance. Second, in most cases, the decrease in recommendation performance is greater when the contrastive learning module is removed compared to when the generated positive samples are excluded from the training data. This may be because both the generated positive and negative samples are used in the contrastive learning loss function, while the increased training data only involves the generated positive samples. Additionally, we observe a sharp decline in recommendation accuracy on the Sports dataset when the contrastive learning module is removed, indicating that the negative samples we generated effectively combined with contrastive learning to enhance recommendation performance on this dataset.

In summary, the results of the ablation study show that our proposed intent insertion data augmentation method plays a significant role in improving the performance of the recommendation system.

\subsection{Hyperparameter Sensitivity Analysis}

\begin{figure}[htbp]
\centering
\includegraphics[width=1.0\columnwidth]{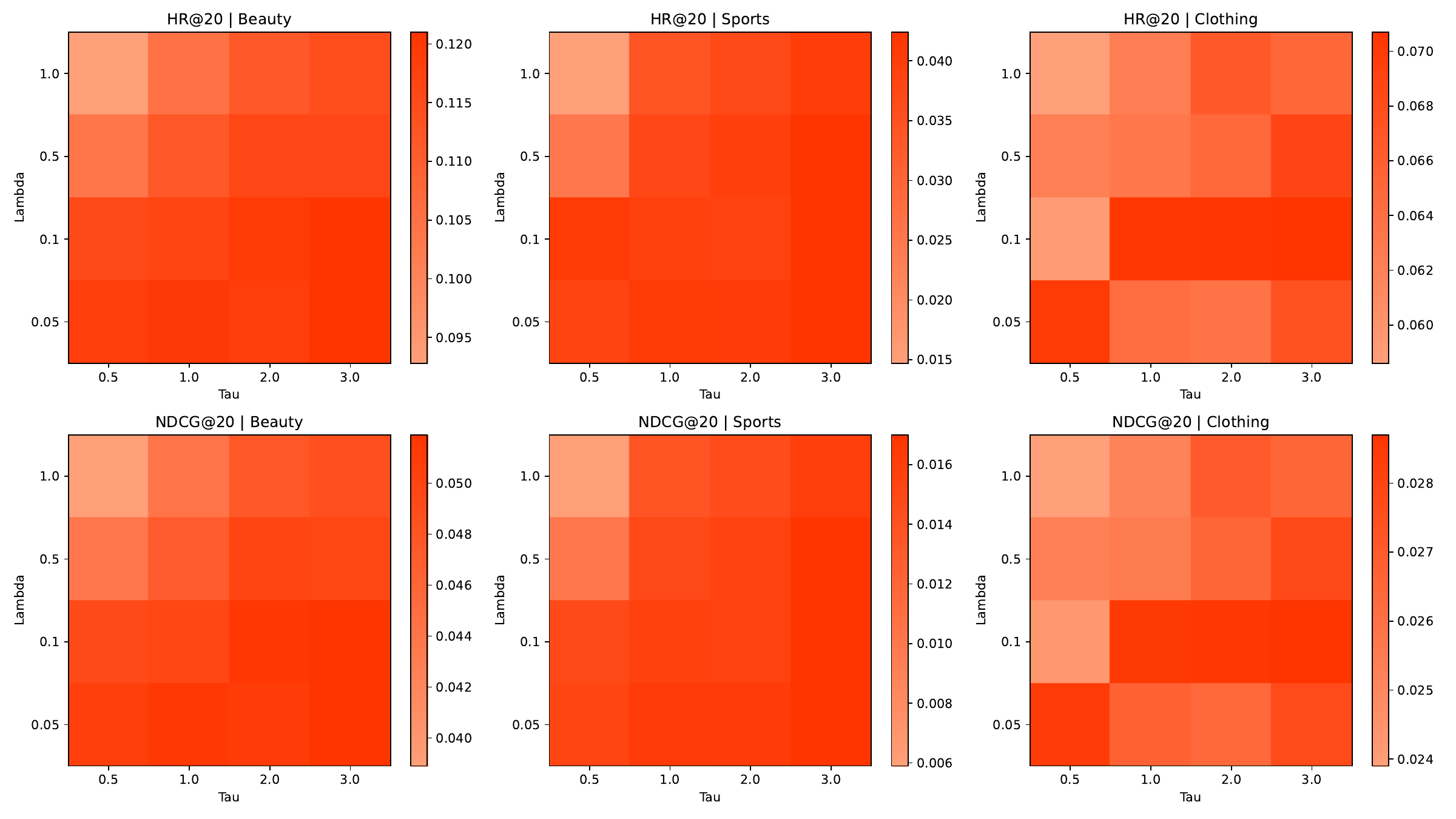}
\caption{Study on Contrastive Learning Hyperparameters of the IESRec Model}
\label{fig:SRec_hyper_cl}
\end{figure}

In our IESRec model, two important hyperparameters are used in the contrastive learning module: \(\lambda\) and \(\tau\). \(\lambda\) represents the importance of the contrastive loss function in the overall loss function, and \(\tau\) is the temperature coefficient in the contrastive loss function. By adjusting these two hyperparameters, we can influence the role of contrastive learning during model training, thereby optimizing the model's recommendation performance. Figure \ref{fig:SRec_hyper_cl} shows the performance of the IESRec model in terms of HR@20 and NDCG@20 on the Beauty, Clothing, and Sports datasets under different values of \(\lambda\) and \(\tau\).

Effect of hyperparameter \(\lambda\): In the Beauty and Sports datasets, as \(\lambda\) decreases, the model's performance in both HR and NDCG metrics initially improves and then stabilizes. Specifically, when \(\lambda\) is small, the weight of the contrastive loss is lower, and the model mainly relies on the original loss function for optimization. Proper integration of contrastive learning helps to improve model performance. As \(\lambda\) increases and the weight of the contrastive loss grows, the recommendation performance declines. Once \(\lambda\) reaches a certain range, model performance tends to stabilize. This suggests that for these datasets, the contrastive loss does not need to be heavily weighted to achieve optimal results. In the Clothing dataset, changes in \(\lambda\) have a relatively greater impact on model performance. Overall, as \(\lambda\) decreases, both HR and NDCG metrics improve to varying degrees, and the effect is strongly related to the value of the other hyperparameter. This indicates that in the Clothing dataset, the weight of the contrastive loss has a more uncertain effect on model performance, and both hyperparameters jointly influence the final recommendation results.

In the Beauty and Sports datasets, the effect of \(\tau\) on model performance is fairly consistent. As \(\tau\) increases, both the HR and NDCG metrics improve. A \(\tau\) value that is too small may lead to an overly strong effect from contrastive learning, causing the model to perform poorly in capturing subtle differences. Thus, it is necessary to find a balance. In the Clothing dataset, the effect of \(\tau\) is more pronounced and is strongly related to the value of \(\lambda\). When \(\lambda\) is small, increasing \(\tau\) results in slight performance improvement, but the overall performance does not reach its best. When \(\lambda\) is larger, the impact of \(\tau\) on recommendation performance becomes inconsistent. This suggests that in the Clothing dataset, the effects of these two hyperparameters are interdependent, and finding a reasonable combination of both is key to achieving optimal results.

\section{Conclusion}

In this chapter, we propose an intent-enhanced sequential recommendation algorithm, IESRec, aimed at addressing issues such as increased data noise and blurred user intent, which are common in current data augmentation methods for sequential recommendation algorithms. IESRec generates positive and negative samples using the concept of intent insertion. First, we generate positive samples by inserting artificial user intent sequences into non-terminal positions in the sequence data. These inserted sequences are derived from the user's historical behavior data and maintain intent continuity with the preceding and following subsequences at the insertion point. Second, we generate negative samples by inserting similar artificial sequences at the end of the sequence data. The generated positive samples are mixed into the training set for training. Additionally, both positive and negative samples are used to construct the contrastive loss function, which is jointly trained with the main task to enhance the overall model performance.

Experimental results on the Beauty, Clothing, and Sports datasets show that the IESRec model significantly outperforms the baseline models. In the ablation experiments, we demonstrated the effectiveness of our data augmentation method. Furthermore, by tuning the hyperparameters in the contrastive learning process, we further improved model performance, verifying the effectiveness of contrastive learning in sequential recommendation tasks. In summary, the IESRec model, by designing a novel intent-sequence-insertion-based data augmentation method, significantly improves the accuracy and stability of sequential recommendation.

\bibliographystyle{alpha}
\bibliography{sample}

\end{document}